\documentclass[dvips]{arxstspdf}
\usepackage{dcolumn}
\usepackage{graphicx,flushend}

\volume{24}
\issue{4}
\pubyear{2009}
\firstpage{547}
\lastpage{560}
\doi{10.1214/09-STS286}
\setattribute{copyright}{owner}{In the Public Domain, 2009}

\makeatletter

\newcolumntype{d}[1]{D{.}{.}{#1}}

\makeatother

\begin{document}
\begin{frontmatter}

\title{On Combining Data From Genome-Wide Association Studies to
Discover Disease-Associated SNPs}
\runtitle{On Combining Genome-Wide Association Studies}

\begin{aug}
\author[a]{\fnms{Ruth M.} \snm{Pfeiffer}\ead[label=e1]{pfeiffer@mail.nih.gov}\corref{}},
\author[b]{\fnms{Mitchell H.} \snm{Gail}\ead[label=e2]{gailm@mail.nih.gov}}
\and
\author[c]{\fnms{David} \snm{Pee}\ead[label=e3]{PeeD@imsweb.com}}
\runauthor{R. M. Pfeiffer, M. H. Gail and D. Pee}

\affiliation{Biostatistics Branch, Division of Cancer Epidemiology and Genetics
National Cancer Institute}

\address[a]{Ruth M. Pfeiffer is Senior Investigator, Biostatistics Branch,
Division of Cancer Epidemiology and Genetics, NCI, NIH, 6120 Executive Blvd,
Rockville, MD 20852, USA \printead{e1}.}
\address[b]{Mitchell H. Gail is Senior Investigator, Biostatistics Branch,
Division of Cancer Epidemiology and Genetics, NCI, NIH, 6120 Executive Blvd,
Rockville, MD 20852, USA \printead{e2}.}
\address[c]{David Pee is Senior Statistician, Information Management
Services, Inc., Silver Spring, MD 20904, USA \printead{e3}.}

\end{aug}

%
\begin{abstract}
Combining data from several case-control genome-wide association (GWA)
studies can yield
greater efficiency for detecting associations of disease with single nucleotide
polymorphisms (SNPs) than separate analyses of the component studies. We
compared several
procedures to combine GWA study data both in terms of the power to
detect a disease-associated
SNP while controlling the genome-wide significance level, and in terms
of the detection
probability ($\mathit{DP}$). The $\mathit{DP}$ is the probability that a particular
disease-associated
SNP will be among the $T$ most promising SNPs selected on the basis of
low $p$-values.
We studied both fixed effects and random effects models in which
associations varied
across studies. In settings of practical relevance, meta-analytic
approaches that focus
on a single degree of freedom had higher power and $\mathit{DP}$ than global
tests such as summing
chi-square test-statistics across studies, Fisher's combination of
$p$-values, and forming
a combined list of the best SNPs from within each study.
\end{abstract}

%

\end{frontmatter}

\section{Introduction}

Case-control genome-wide association (GWA) studies are used to detect
associations of disease with genetic markers (alleles of single
nucleotide polymorphisms or SNPs) across the genome by comparing
\mbox{individuals} with disease (cases) to disease-free individuals
(controls). A widely accepted approach for identifying and confirming
an association is to conduct an initial discovery study to detect
promising SNPs and then to validate the associations in data from
independent studies, as, for example, in Easton et al. (\citeyear{2007east}). Both
power calculations (e.g., Skol et al., \citeyear{2007skol}) and calculations of the
probability of detecting disease-associated SNPs (Gail et al., \citeyear{2008agail1})
indicate that large numbers of cases and controls are needed for a
successful discovery study if one is interested in common alleles with
small odds ratios (e.g., odds ratio per allele $=$ 1.2), such as have
been found in GWA studies for breast (Easton et al., \citeyear{2007east}) and prostate
(Yeager et al., \citeyear{2007yeager}) cancer. A recent study of diabetes (Zeggini et
al., \citeyear{2008zegg}) illustrated that combining data from several studies could
improve discovery efforts, compared to the separate analyses of the
component studies. In some diseases, such as thyroid cancer or
amyotropic lateral sclerosis (ALS), it is not possible to accrue large
numbers of cases and controls in a single region or study center; in
this context, data will need to be combined for successful discovery.
In this paper we compare several approaches to using data from several
smaller GWA studies to discover promising disease-associated SNPs that
require further validation studies.

We compare procedures to combine data from ge\-nome-wide association
studies both in terms of the power to detect a disease-associated SNP
while controlling the experiment-wide (including genome-wide)
significance level, and in terms of the detection probability. The
detection probability is the probability that a particular
disease-associated SNP will be among the $T$ most promising SNPs selected
on the basis of low $p$-values (or high chi-square tests).

In Section~\ref{sec2} we describe models for disease association, including a
fixed effects model that assigns the same log-odds ratio to each
disease SNP and a random effects model that allows this log-odds ratio
to vary across studies. In Section~\ref{sec3} we review the concept of detection
probability for a single GWA study and extend the concept for several
procedures for combining data from $S$ case-control studies.
We also define and compute power for these procedures, while
controlling the experiment-wide significance level (Section~\ref{sec4}).
 Section~\ref{sec5} contains numerical results to compare procedures with respect to
detection probability and power. Some conclusions are given in Section~\ref{sec6}.

\section{Data and Models}\label{sec2}

We assume that genotypes for $N$ SNPs from the same genotyping platform
are available for\break case-control studies $s=1,\ldots,S$. In this paper
we let $N=500{,}000$. Study $s$ includes $n_s$ cases and $n_s$ controls.
Let $X_i =0,1$ or $2$ be the number of minor alleles at locus $i$ for
$i=1, \ldots, N$,
and let $Y=1$ for diseased and $0$ for nondiseased subjects. Suppose
SNPs $1, \ldots, M$ are associated with disease, while SNPs $M+1,
\ldots, N$ are not, resulting in the model for disease
\begin{equation}\label{popmodel}
\quad\ \ \operatorname{logit}\{P_s(Y = 1| X_{1}, \ldots, X_N)\} = \mu_s+ \sum
_{i=1}^M \beta
_i^s X_{i}.
\end{equation}
Thus, we assume that the log-odds ratios for the nondisease-associated
SNPs are equal to zero. In numerical studies in Section~\ref{sec5}, we assume
that all disease-associated SNPs have the same log-odds ratio within a
study, $\beta_i^s= \beta^s$ for $i=1, \ldots,M$ and for $s=1,\ldots
,S$. We model variation of $\beta^s$ among studies in two ways. In the
fixed effects model we set $\beta^s=\beta$ for $s=1,\ldots,S$, as
might happen if the cases and controls for the $S$ studies were sampled
from the same homogeneous population.
Under a random effects model, the log-odds ratios for the disease
related SNPs are independent normal variables, $\beta^s \sim N(\beta,
\tau^2),$ $s=1,\ldots,S$. As tagging SNPs are typically only markers
in linkage disequilibrium (LD) with the true causal disease SNPs, this
model captures the impact of variation in LD patterns on $\beta^s$
across study populations.

We have assumed that log-odds ratios are strictly zero for the $N-M$
nondisease-associated SNPs. This ``strong null hypothesis'' is
plausible because, if there is no nearby disease SNP, then no amount of
LD among nearby SNPs can induce an association between a marker SNP and disease.

\section{Methods to Compute Detection Probability from Combined
Studies}\label{sec3}

\subsection{Review of Detection Probability for a Single Case-Control
GWA Study}\label{sec3.1}

In a single GWA study, if disease is rare and the SNP scores $X_i$ are
independent in the source population,
\begin{eqnarray}\label{model0}
 \operatorname{logit}\{ P(Y=1|X_i) \} = \mu^*+ \beta_i X_{i},
\\ \eqntext{i=1,\ldots,N,}
\end{eqnarray}
in the case-control population (Gail et al., \citeyear{2008agail1}). In (\ref{model0})
$\mu^*= \mu+ \log\{E(\exp(\sum_{k \neq i}^M \beta_k X_k)\} + \log
(\pi_1/\pi_0)$, where $\pi_1$ is the proportion of cases in the
source population that are in the case-control study, and $\pi_0$ is
the analogous proportion for controls. $E$ is the expectation operator.

The null hypothesis of no association for the $i$th SNP, $H_0\dvtx \beta
_i=0$, can be tested using the Wald statistic for a trend in risk with
the number of minor alleles, $W_i= \hat\beta_i^2/ \operatorname{var}(\hat\beta
_i)$, where $\hat\beta_i$ denotes the maximum likelihood estimate for
model (\ref{model0}) and its variance $\operatorname{var}(\hat\beta_i)$ is computed
under the retrospective sampling
(Gail et al., \citeyear{2008agail1}). Alternatively, one could use the score test for
trend (Armitage, \citeyear{1955arm}).
Under the null hypotheses of no association, both the Wald and the
score test have one degree of freedom chi-square ($\chi^2_1$)
distributions. These tests correspond to additive (or codominant)
genotype scores (Sasieni, \citeyear{1997sas}) and yield the same value whether the
major or minor allele is positively associated with disease (Devlin and
Roeder, \citeyear{1999dev}; Pfeiffer and Gail, \citeyear{2003pfei}). Moreover, under the rare
disease assumption, the $W_i$ are independent, which facilitates the
calculation of detection probability (Gail et al., \citeyear{2008agail1}).

A particular SNP, for example, SNP $k$, is \textit{T-selected} or simply
\textit{selected} if its associated Wald statistic (or $p$-value) is among
the top $T$ test statistic values (or $T$ lowest $p$-values), that is,
$\operatorname{rank}(W_k)>N-T$. The probability that a particular {\it
disease-associated} SNP, for example, SNP $i$, is $T$-selected is the
\textit{detection probability} ($\mathit{DP}$), that is, $\mathit{DP}=P(\operatorname
{rank}(W_i)>N-T)$. The
\textit{proportion positive} ($\mathit{PP}$) is the fraction of selected SNPs that
are true disease-associated SNPs.

\subsection{Combined List of SNPs}\label{sec3.2}

Here, each of the $S$ studies is analyzed separately. The Wald test
statistics $W^s_j,  j=1,\ldots,N$, based on model (\ref{model0})
are ranked within study $s$, for $s=1, \ldots, S$, and in each study
the top $T/S$ SNPs are selected. We then create a ``combined list'' of
the union of the sets of $T/S$ SNPs selected from each study. We let
$T^c$ be the number of distinct SNPs that are $T/S$ selected in at
least one of the $S$ studies.
$T^c$ is not a fixed number, but a random variable, with $T/S \leq T^c
\leq T$, depending on the amount of overlap among the top $T/S$ SNPs
from the $S$ studies.

As the $S$ studies are independent, the probability that disease SNP
$i$ is $T/S$ selected in $k$ out of $S$ studies is given by
\begin{eqnarray*}
&&P(\mbox{SNP $i$ $T/S$-selected in $k$ studies})
\\
&&\quad = \sum_{A_k} \prod
_{l \in A_k} \mathit{DP}^l_{i} \prod_{l \notin A_k} (1-\mathit{DP}_i^l),
\end{eqnarray*}
where $\mathit{DP}_i^s$ denotes the detection probability for the $i$th disease
SNP in study $s$, that is, $\mathit{DP}^s_i=\break P(\operatorname{rank}(W^s_i)> N-T/S)$,
and the sum is over all $S!/k!(S-k)!$ ways of selecting the set of $k$
indices, $A_k$, from the set $\{1, \ldots, S\}$.
$\mathit{DP}_i^s$ is computed either under a fixed effects or random effects
model for the log-odds ratios of the
disease-associated SNPs. If the studies are exchangeable and $\mathit{DP}^s_i
=\mathit{DP}_i$ for all $s$, $P$(SNP $i$ $T/S$-selected in $k$
studies) simplifies to a binomial probability and the expected number
of studies that $T/S$-select the $i$th disease SNP is $S(\mathit{DP}_i).$

The \textit{combined detection probability}, namely, the probability that
the $i$th disease SNP is $T/S$ selected in at least one of the $S$
studies, is
\begin{equation}\label{listDP}
\mathit{DP}_i= 1-\prod_{s=1}^S(1-\mathit{DP}^s_i).
\end{equation}

For special settings, analytic expressions for $\mathit{DP}^i_s$ given in Gail
et al. (\citeyear{2008agail1}) can be
used in (\ref{listDP}) to approximate
$\mathit{DP}_i$. When all the studies have the same sample size and when there
is only a single disease-associated SNP, $M=1$, that has the same fixed
log-odds ratio $\beta$ in (\ref{model0}) for each individual study,
\begin{equation}\label{DP1}
\mathit{DP} \approx 1-[F_{H_1}(\chi^2_{1,1- T/SN})]^{S}.
\end{equation}
In expression (\ref{DP1}) $\chi^2_{1,1- T/SN}$ denotes the $1-T/SN$
quantile of a central $\chi^2_1$ distribution, and $F_{H_1}$ denotes a
noncentral chi-square distribution $\chi^2_1(\delta)$ with
non-centrality $\delta=\beta^2 /\sigma^2_1$, where $\sigma^2_1$ is
given in equation (\ref{var0}) in the \hyperref[app]{Appendix}.

The expected proportion of positive findings out of the $T^c$ SNPs is
approximately
\[
\mathit{PP} = E \biggl\{ \frac{ \sum_{i=1}^M \mathit{DP}_i}{T^c} \biggr\} \approx\frac
{ \sum_{i=1}^M \mathit{DP}_i}{T},
\]
because, as demonstrated in simulations (Section \ref{sec5.1}),
there is very little overlap among selected SNPs across studies and,
therefore, $T^c$ is usually close to $T$.

\subsection{Pooled Individual Level Data}\label{sec3.3}

We show in Section \ref{s:rankmeta} that a meta-analytic approach has
equivalent efficiency to pooling individual level data. Therefore, in
numerical studies below we only use the meta-analytic approach.
Nonetheless, it is instructive to outline an analysis of individual
level data from $S$ studies with the following fixed effects model.

We assume that the log-odds parameter, $\beta_i$, for disease SNP $i$
is the same in all studies, leading to
\begin{eqnarray}\label{model1}
\operatorname{logit}(p_{si})&=& \operatorname{logit}\bigl(P_s(Y=1|X_i)\bigr)\nonumber
\\[-8pt]\\[-8pt]
&=&
\mu_s^*+ \beta_i X_{i}, \quad s=1,\ldots
, S,\nonumber
\end{eqnarray}
where $\mu_s^*$ denotes the study-specific intercept that accommodates
differences in disease prevalence and differences in sampling fractions
among the different studies. The Wald statistic for the $i$th SNP is
computed by first finding the estimate $\hat\beta_i$ that maximizes
the likelihood
\begin{equation}\label{likelihood1}
\quad L(\beta_i,\mu_1^*, \ldots, \mu_S^*) = \prod_s \prod_j
p_{sj}^{Y_{sj}}(1 - p_{sj})^{1-Y_{sj}}.
\end{equation}
The information matrix to compute the variance of $\hat\beta_i$
depends on the study specific intercepts $\mu_s$. An expression for
$\operatorname{var}(\hat\beta_i)=\sigma^2_{Si}$ is provided in equation (\ref
{var1}) in the \hyperref[app]{Appendix}. The corresponding Wald test statistic $
W_i=\hat\beta_i^2 / \sigma^2_{Si}$ has a central $\chi^2_1$
distribution if $\beta_i=0$
and a noncentral $\chi^2_1(\delta)$ distribution with $\delta=\beta
^2_i/\sigma^2_{Si}$ otherwise.

Selection of the top $T$ SNPs is based on ranking the Wald statistics
$W_i$, $i=1, \ldots, N$, computed from model (\ref{model1}).
If $M=1, n_s=n$, and $\beta_i^s=\beta_i$ for $s=1,\ldots, S$, then,
following Gail et al. (\citeyear{2008agail1}),
\begin{equation}\label{approx}
\mathit{DP} \approx1-F_{H_1}(\chi^2_{1,1- T/N}),
\end{equation}
where $F_{H_1}$ is a noncentral $\chi^2_1(\delta)$ distribution with
noncentrality parameter $\delta=\beta^2_i/\sigma^2_{Si}$.

\subsection{Meta-Analytic Approaches}\label{s:rankmeta}

We first estimate study-specific log-odds ratios $\hat\beta_i^s$ for
the $i$th SNP, $i=1,\ldots, N$, by fitting model (\ref{model0})
separately to each SNP for each study and then combine study specific
maximum likelihood estimates $\hat\beta_i^s$ to obtain an overall
estimate of disease association for the ith SNP. This can be done using
a fixed effects model (Mantel and Haenszel, \citeyear{1959mantel}; Yusuf et al., \citeyear{1985yusuf})
or a random effects model (DerSimonian and Laird, \citeyear{1986der}) for disease SNPs.

For the fixed effects model, the combined SNP specific estimate is
\begin{equation}\label{model3}
\hat\beta^F_i= \sum_{s=1}^S \hat\beta_i^s \hat w_i^s,
\end{equation}
where $\hat w_i^s = (1/ \hat\sigma^2_{is}) (\sum_{k=1}^S 1/ \hat
\sigma^2_{ik})^{-1}$.
Under the null hypothesis of no association, $ \hat\beta^F_i$ has an
asymptotic normal distribution with mean zero and variance\break $\operatorname{var}(\hat
\beta^F_i)=(\sum_{k=1}^S 1/ \sigma^2_{ki})^{-1}$. As shown in the\break
\hyperref[app]{Appendix}, $\operatorname{var}(\hat\beta^F_i)= \sigma^2_{Si}$, the variance of
the\break
maximum-likelihood estimate based on model (\ref{model1}). Thus, the
two approaches are equally efficient under the fixed effects model and
in Section~\ref{sec5} we only study the meta-analytic approach.

Under a random effects model (DerSimonian and Laird, \citeyear{1986der}), estimates
$\hat\beta^s_i$ are assumed to follow a linear model, $\hat\beta
^s_i= \beta_i^s + \epsilon_i^s$, where $\beta_i^s$ is a normal
variate with mean $\beta_i$ and variance $\tau^2_i$, the $\epsilon
_i^s$ are normally distributed with mean zero and variance $\sigma
^2_{is}$, and $\beta^s_i$ and $\epsilon_i^s$ are independent. Thus,
under the random effects model $\operatorname{var}(\hat\beta_i^s)= \sigma^2_{is} +
\tau^2_i$. Note that this model is equivalent to the random effects
model for disease SNPs in Section~\ref{sec2} and that
$E(\hat\beta^s_i)^2=\beta^2 + \sigma^2_{is} + \tau^2_i$, which can
be large even when $\beta=0$. The strong null hypothesis for
nondisease-associated SNPs, however, corresponds to a fixed effects
model with $\beta_i=0$ or, equivalently, to a degenerate random
effects model with $\beta_i^s=0$ and $\tau_i^2=0$. Replacing the
$\sigma^2_{is}$ by their estimates reported in the individual studies,
we have (DerSimonian and Laird, \citeyear{1986der})
\[
\hat\tau^2_i= \max\biggl\{ 0,\frac{\sum_s u_{is}(\hat\beta^s_i-
\hat\beta^F_i)^2-(S-1)}{\sum_s u_{is} - \sum_s u_{is}^2/\sum_s
u_{is}} \biggr\},
\]
where $u_{is} = 1/ \sigma^2_{is}$ and $\hat\beta^F_i$ is given by
(\ref{model3}).
The random effects meta-analytic estimate of the association of the ith
SNP with disease is then given by
\begin{equation}\label{model4}
\hat\beta^R_i= \sum_{s=1}^S \hat\beta_i^s \hat v_{is},
\end{equation}
where $\hat v_{is} = (\hat\tau_i^2+ \hat\sigma^2_{is})^{-1}/\{ \sum
_{k=1}^S (\hat\tau_i^2+ \hat\sigma^2_{ik})^{-1}\}$. The variance of
$\hat\beta^R_i$ is therefore approximated by $\operatorname{var}(\hat\beta^R_i) =
1 /\{ \sum_{k=1}^S (\hat\tau_i^2+ \hat\sigma^2_{ik})^{-1}\}$.

In order for the between study variance $\tau^2_i$ to be reliably
estimated, the number of studies $S$ cannot be too small. For the fixed
effects model, $\hat\beta^F_i$ becomes asymptotically normal as $n_S$
increase. For the random effects model, $\hat\beta^R_i$ becomes
asymptotically normal as $S$ increases.

The detection probabilities are computed by ranking the Wald statistics
$ W_i^F=(\hat\beta^F_i)^2/ \sigma^2_{Si}$, for the fixed effects
meta-analytic approach, or $ W_i^R=\break(\hat\beta^R_i)^2 / \operatorname{var}(\hat\beta
^R_i)$ for the random effects meta-analytic approach.

\subsection{Sums of Test Statistics and Fisher Combination of
\textit{p}-Values}\label{s:rankfisher}

Let $W_i^s$ denote the Wald test statistics for SNP $i$ in study $s$
obtained from fitting (\ref{model0}) to the study-specific data.
The combined test statistic is
\begin{equation}
\label{wald}
W_i= \sum_{s=1}^S W_i^s,
\end{equation}
which, for the nondisease-associated SNPs, has a central $\chi^2_S$
distribution.
For the disease-associated SNPs, and conditional on $\beta^s_i$, $W_i$
has a noncentral $\chi^2_S(\delta)$ distribution with noncentrality
parameter $\delta=\sum_{s=1}^S (\beta^s_i)^2/\sigma^2_{is}$.
For $M=1$ and $\beta_1^s=\beta$, the detection probability is well
approximated by (\ref{approx}). For this special case $\delta=\beta
^2 S /\sigma^2_{1},$ where $\sigma^2_{1}$ is specified in the
\hyperref[app]{Appendix}
formula (\ref{var0}).

Instead of combining the Wald statistics, one can combine $p$-values
$p^s_i$ across studies,
through $ p^c_i= \prod_{s=1}^S p_i^s$ (Fisher, \citeyear{1932Fisher}), and rank SNPs
based on $p^c_i$.
Under the null hypothesis, $-2 \log p_i^c = -2 \sum_{i=1}^S \log
p_i^s$ has a central $\chi^2_{2S}$ distribution.
Numerous other combinations of $p$-values have been proposed and studied
(Loughin, \citeyear{2004lough}). We therefore also
assessed the performance of the Liptak--Stouffer combination of
$p$-values, given by $LS= \sum_{i=1}^S \Phi^{-1}(1- p_i^s)/\sqrt{S}$,
that has a normal distribution with mean zero and variance one under
the null hypothesis (Liptak, \citeyear{1958LIPTAK}).

\section{Power of Various Approaches to Combining GWA Studies}\label{sec4}

Except for the Fisher and Liptak--Stouffer methods of combining
$p$-values, we computed the statistical power of the approaches to
combining data presented in Sections \ref{sec3.2}--\ref{s:rankfisher} analytically based on
asymptotic theory, and also tested analytical results in simulations.
The power is
the probability that the test statistic for a given SNP will fall into
the predetermined critical region that is chosen to control
the significance level for multiple testing of the $N$ genotypes and
$S$ studies. In contrast to the ranking procedures for detection
probabilities, the power for any particular SNP does not depend on the
test statistic for any other SNP. We therefore usually omit the SNP
index in what follows.
The rejection region is chosen based on the strong null hypothesis that
the log-odds ratios for the nondisease-associated SNPs are always equal
to zero, regardless of the model that gives rise to the effects for the
disease-associated SNPs.

We set $\alpha=0.05/N=10^{-7}$ to account for multiple testing.
Further control of multiplicity for $S$ is described below.

\subsection{Combine Lists of Significant SNPs from Each Study}\label{sec4.1}

As in Section~\ref{sec3.2}, we compute study-specific Wald statistics $W^s_j$,
$j=1,\ldots,N$, $s=1,\ldots, S,$ based on model (\ref{model0}).
We determine significance based on\break whether $W^s_j$ exceeds the
significance threshold\break $\chi^2_{1,1-\alpha}$, the $1-\alpha$
quantile of a $\chi^2_{1}$ distribution. As we are combining results
from $S$ studies, we replace $\alpha$ by $\alpha/S$ to control the
experimentwise error at $0.05$. An exact calculation replaces $\alpha$
by $\alpha^*=1-(1-\alpha)^{1/S}$, but for small $\alpha$ this
$\alpha^*$ is very nearly $\alpha/S$.

The power of the combined list approach under an alternative $H_1$ is thus
\begin{eqnarray}
\label{listDPpower}
&&P_{H_1}(W^s > \chi^2_{1,1-\alpha/S} \mbox{ in at least one study})\nonumber
\\[-8pt]\\[-8pt]
&&\quad =
1-\prod_s P_{H_1}(W^s \leq\chi^2_{1,1-\alpha/S}).\nonumber
\end{eqnarray}
When all the disease-associated SNPs for the different studies have the
same fixed effect, $\beta^s=\beta$, $P_{H_1}$ is generated by a $\chi
^2_1(\delta)$ distribution with $\delta=\beta^2 / \sigma^{2}_{1s}$,
where $\sigma^{2}_{1s}$ is given in equation (\ref{var0}) in the
\hyperref[app]{Appendix}. When all the studies have the same sample size, then (\ref
{listDPpower}) reduces to $1-[F_{H_1}(\chi^2_{1,1-\alpha/S})]^{S}$,
which is equivalent to (\ref{DP1}) with $T=\alpha N$.

To obtain the power when the log-odds ratios of the disease-associated
SNPs arise from the random effects model, $\beta^s \sim N(\beta, \tau
^2), s=1,\ldots, S$, we integrate (\ref{listDPpower}) over the
distribution of the independent study specific $\beta^s$ parameters to obtain
\begin{eqnarray*}
&&P_{H_1}(W^s > \chi^2_{1,1-\alpha/S} \mbox{ in at least one study})
\\
&&\quad=
1-\prod_s \int_{\beta^s} P_{H_1}(W^s \leq\chi^2_{1,1-\alpha/S};
\beta^s)\, dF(\beta^s),
\end{eqnarray*}
where $F$ denotes the normal distribution with mean $\beta$ and
variance $\tau^2$.

\subsection{Meta-Analytic Approaches}\label{sec4.2}

\subsubsection*{Fixed effects meta-analytic approach}
Based on\break asymptotic normal theory, the power
for the test statistic $ W^F=(\hat\beta^F)^2/ \operatorname{var}(\beta^F)$ is
\begin{equation}\label{powermeta1}
P_{H_1}(W^F > \chi^2_{1,1-\alpha} ).
\end{equation}
Under the fixed effects model for the disease-associated SNPs,
$P_{H_1}$ is generated by a $\chi^2_1(\delta)$ distribution with $
\delta= (\beta^F)^2/ \sigma^2_{S}$, where $\beta^F= \sum_{s=1}^S
\beta^s w^s,$ $ w^s =
(1/\break \sigma^2_{s}) (\sum_{k=1}^S 1/ \sigma^2_{k})^{-1}$ and
$\sigma^2_{S}$ is given in the \hyperref[app]{Appendix} equation (\ref{var1}). The power
under the random effects model for disease-associated SNPs is obtained
by integrating equation (\ref{powermeta1}) over the distribution of
$\beta^s$, namely, $ P_{H_1}(W^F > \chi^2_{1,1-\alpha} ) = \int
_{\beta^1} \cdots\int_{\beta^S} P_{H_1}(W^F > \chi^2_{1,1-\alpha
}; \beta^1, \ldots, \beta^S)\, dF(\beta^1)\cdots\, dF(\beta^S)$.

\subsubsection*{Random effects meta-analytic approach}

The use of asymptotic normal theory for the random effects
meta-analytic approach when there are few studies is problematic, as
the type I error rate can be substantially inflated (Follmann and
Proschan, \citeyear{1999foll}). Follmann and Proschan therefore suggest using a
$t_{S-1}$ reference distribution rather than a standard normal
distribution. Using the $t$-approximation, the power of the random
effects meta-analytic approach is
\begin{equation}
\label{powerrandom}
P_{H_1}(W^R > F_{1,S-1,1-\alpha} ),
\end{equation}
where $P_{H_1}$ is generated by a noncentral $F_{1,S-1}$ distribution,
with noncentrality parameter $ \delta=(\beta^R)^2 /\sigma^2$,
and $F_{1,S-1,\alpha}$ is the $1-\alpha$ quantile of a central
$F_{1,S-1}$ distribution.
However, under the strong null hypothesis that the log-odds ratio
parameters for the\break nondisease-associated SNPs are strictly zero and do
not vary across studies, one can replace the\break $F_{1,S-1,1-\alpha}$
cutoff value in (\ref{powerrandom}) by
$\chi^2_{1,1-\alpha}$, as for the fixed effects meta-analytic
approach. In simulations we study the power for the random effects
meta-analytic approach using both cutoff values for the test statistic.

The power under the random effects model is obtained by integrating
equation (\ref{powerrandom})
or $P_{H_1}(W^R> \chi^2_{1,1-\alpha})$, over the random effects
distribution of the $\beta^s,$ similar to the fixed effects
meta-analytic approach given above.

\subsection{Power of the Sum of Test Statistics}\label{sec4.3}

The power for the test statistic $ W= \sum_{s=1}^S W^s$ is given by
\begin{equation}
P_{H_1}(W > \chi^2_{S,1-\alpha}),
\end{equation}
where $P_{H_1}$ is generated by a $\chi^2_S(\delta)$ distribution
with $\delta=\sum_{s=1}^S  (\beta^s)^2/\sigma_s^2$.

We do not compute the power for Fisher's\break $-2 \sum_{s=1}^S \log p^s$ or
the Liptak--Stouffer combination of $p$-values
analytically, because the distribution
of the $S$ $p$-values $p_1, \ldots, p_S$ cannot be obtained in a
manageable form under the alternative.

\section{Simulations} \label{sec5}

\subsection{Simulation Methods to Estimate the Detection Probability, DP} \label{sec5.1}

We used the methods in Gail et al. (\citeyear{2008agail1}) for a single study to
simulate data separately from each of the
case-control studies, $s=1,\ldots, S$.
At each SNP $i=1,2, \ldots,N$, we randomly and independently selected
a minor allele frequency, $\eta_i$, from the
distribution of minor allele frequencies in CGEMS\break
(\url{https://caintegrator.nci.nih.gov/cgems/}), as described in Gail et al.
(\citeyear{2008agail1}). In each replicate of the simulations described below, minor
allele frequencies were re-assigned to each SNP in this way. We assumed
that the $N$ genotypes were statistically independent in the source
population, the disease is rare and the Hardy--Weinberg equilibrium
holds at each locus. Given $\beta_i$, we sampled $\hat\beta_i$ from
$N(\beta_i, \sigma^2_i(\beta_i))$ independently for each $i=1,\ldots
,N$ to generate realizations of the Wald statistics rapidly in GAUSS
(Aptec Systems, 2005). The Wald statistics were computed as $W_i=\hat
\beta_i^2/\sigma^2_i(\beta_i)$, which has the same asymptotic
distribution as $\hat\beta^2_i / \hat\sigma^2_i(\beta_i)$.

For each disease model and parameter setting we generated $\mathit{NSIM}=1000$
independent simulations.
Under either the fixed or random effects disease model, and conditional
on $\eta_i$ and $\beta^s$, we computed $\sigma_s^2= \operatorname{var}(\hat\beta
^s)$ and then drew $\hat\beta^s$ from $N(\beta^s, \sigma_s^2)$. The
study-specific estimates were then used in the procedures
in Sections \ref{sec3.2}, \ref{s:rankmeta} and \ref
{s:rankfisher} to compute $\mathit{DP}$.

Define $I(m, \mathit{ISIM},T)=1$ if the rank of the corresponding test statistic
falls into the top $T$ ranks of the N ranked values of the test
statistics in simulation $\mathit{ISIM}$, and 0 otherwise. The detection
probability for each approach is then estimated by
\[
\widehat{\mathit{DP}} = \mathit{NSIM}^{-1} M^{-1} \sum_{\mathit{ISIM}=1}^{\mathit{NSIM}} \sum_{m=1}^M I(m, \mathit{ISIM},T).
\]
$\mathit{PP}$ was estimated from $\widehat{\mathit{PP}} = (\widehat{\mathit{DP}}) M/T$. For the combining
lists approach, we modified these formulas to take into account
variation in $T^c$. Letting $I(m, \mathit{ISIM},T/S)=1$ if the disease SNP
is\break
$T/S$-selected in any study in simulation $\mathit{ISIM}$ and 0 otherwise, we
estimated $\mathit{DP} $ as above with $I(m, \mathit{ISIM},\break T/S)$ in place of $I(m,
\mathit{ISIM},T)$, and we estimated $\mathit{PP}$ from
$\widehat{\mathit{PP}}= \mathit{NSIM}^{-1} \sum_{\mathit{ISIM}} \sum_m I(m, \mathit{ISIM},\break T/S)/T^c(\mathit{ISIM})
$, where $T^c(\mathit{ISIM})$ is the cardinality of the union of the $S$
$T/S$-selected sets of SNPs.

\subsection{Simulations to Estimate Power}\label{sec5.2}

We estimated power by simulations for each of the procedures in Section
\ref{sec4}. We fixed the allele frequency for the disease-associated SNP at
$\eta=0.2673$, the mean allele frequency used in the $\mathit{DP}$
calculations. Estimates $\hat\beta$ were otherwise obtained as in
Section \ref{sec5.1}, but for a single locus.

We used $\mathit{NSIM}=100{,}000$ replicates of outcome data and for each
replicate, each of the test statistics was calculated, and the true
power estimated as the proportion of replicates which were significant
at the experimentwise level
$\alpha= 10^{-7}$.

\begin{table*}[t]
\caption{Detection Probability (DP) and Proportion Positive (PP) in
percent for five methods of combining data from $S$
studies with $n_s$  cases and $n_s$ controls  for   fixed
effects models with $\beta=\log(1.3)$, $N=500{,}000$ SNPs,
 and random allele frequency $\eta$}\label{tab1}
\begin{tabular*}{\tablewidth}{@{\extracolsep{4in minus 4in}}ld{2.2}d{2.2}d{2.2}d{2.2}d{2.2}ccccd{1.3}@{}}
\hline
 \textbf{Method} &     \multicolumn{2}{c}{$\bolds{T=20}$}  &  \multicolumn{2}{c}{$\bolds{T=100}$}   & \multicolumn{2}{c}{$\bolds{T=1000}$}
  & \multicolumn{2}{c}{$\bolds{T=10{,}000}$ }  &
  \multicolumn{2}{c@{}}{$\bolds{T=25{,}000}$}\\
 \ccline{2-3,4-5,6-7,8-9,10-11}
&  \multicolumn{1}{c}{$\bolds{\mathit{DP}}$} & \multicolumn{1}{c}{$\bolds{\mathit{PP}}$} & \multicolumn{1}{c}{$\bolds{\mathit{DP}}$} & \multicolumn{1}{c}{$\bolds{\mathit{PP}}$}
  &      \multicolumn{1}{c}{$\bolds{\mathit{DP}}$} & \multicolumn{1}{c}{$\bolds{\mathit{PP}}$}  & \multicolumn{1}{c}{$\bolds{\mathit{DP}}$} & \multicolumn{1}{c}{$\bolds{\mathit{PP}}$} & \multicolumn{1}{c}{$\bolds{\mathit{DP}}$} & \multicolumn{1}{c}{$\bolds{\mathit{PP}}$}  \\
\hline
 \multicolumn{11}{c}{$S=5, n_s=400,   M=1$ true disease SNP }    \\
  Comb list &      7.20  & 0.36  &   15.70  &0.16  &      38.10& 0.04 &    73.80  &0.01 &   85.30  &0.003 \\
Ave $T^c$   &   \multicolumn{2}{c}{  20.0  }&    \multicolumn{2}{c}{  100.0 }&   \multicolumn{2}{c}{ 999.0 }&
 \multicolumn{2}{c}{ 9919.5}&  \multicolumn{2}{c}{ 24504.0} \\
Meta fixed  &   74.20 & 3.71&      81.50 &  0.82&      91.00&  0.09   &  96.80  &  0.01 & 98.20 & 0.003 \\
Meta random& 74.20 &   3.71 &    81.50&  0.82  &     91.00 & 0.09 &    96.80 &  0.01  & 98.20  & 0.003 \\
$\sum_s W_s$  &    53.90 &  2.70 &      64.70 & 0.65 &       79.20  &  0.08 &    90.30 &   0.01 & 93.90 & 0.004 \\
  $-2 \sum_s \ln(p_s)$   &   58.40 &  2.92 &     66.90 &  0.67  &    80.20 &  0.08 &   90.80 &    0.01 &  94.50  &  0.004
  \\[5pt]
      \multicolumn{11}{c}{$S=5, n_s=400,  M=10$ true disease SNPs }    \\
  Comb list &    7.75 & 3.89  &  16.95  & 21.70 &     41.61   &  0.42 &   74.42   & 0.08 &  85.46  & 0.03 \\
Ave $T^c$   &\multicolumn{2}{c}{ 20.0}&     \multicolumn{2}{c}{ 99.8  }&  \multicolumn{2}{c}{998.0  }&
 \multicolumn{2}{c}{9914.4 }& \multicolumn{2}{c}{24494.6}\\
Meta fixed  &    73.15 &6.58 &   82.45 & 8.25 &       91.87 &0.92 &   97.53 &0.10 &  98.78 &0.040 \\
Meta random&  73.15 &36.58 &      82.45 & 8.25 &      91.87  &0.92 &    97.53  &0.10 & 98.78& 0.040 \\
$\sum_s W_s$  &     53.12& 26.56&    65.23 &  6.52 &    79.70 & 0.80   & 91.11 &0.09 &    94.87    & 0.038 \\
 $-2 \sum_s \ln(p_s)$  &    55.98  &  27.99&     67.47  & 6.75 &      81.27 & 0.81&     91.68 & 0.09&  95.34 &     0.038
 \\[5pt]
  \multicolumn{11}{c}{$S=10, n_s=200,  M=1$ true disease SNP }    \\
  Comb list &    1.10  &   0.06 &  2.50  &  0.04 & 7.30 & 0.02 &    50.70 & 0.01& 68.60 & 0.003   \\
Ave $T^c$ & \multicolumn{2}{c}{20.0  }&   \multicolumn{2}{c}{100.0  }&  \multicolumn{2}{c}{999.1    }&
\multicolumn{2}{c}{ 9910.7 }& \multicolumn{2}{c}{24445.0 } \\
Meta fixed & 73.20 & 3.66 &  80.50 & 0.81 &  90.80&  0.09&    96.40&  0.01 &  98.30     &    0.004\\
Meta random&   73.20  &  3.66 &    80.50 &0.81  &     90.80 & 0.09 &   96.40 & 0.01  &   98.30 &0.004\\
 $\sum_s W_s$  &    39.00 & 1.95 &   50.40 &  0.50 &    68.60 &0.07 &  83.80 &0.01 &  88.90         &  0.004\\
 $-2 \sum_s \ln(p_s)$  &     42.00  &2.10 &      53.00  &0.53 &     69.70  &  0.07 &  84.30  &  0.01 & 89.60  &
 0.004\\[5pt]
 \multicolumn{11}{c}{$S=10, n_s=200,  M=10$ true disease SNPs }    \\
  Comb list &     1.50   & 0.75  &    4.44 &  0.44  &   17.20  & 0.17 & 49.48 &   0.05& 67.64 &  0.03\\
Ave $T^c$ & \multicolumn{2}{c}{ 20.0}&     \multicolumn{2}{c}{100.0  }&  \multicolumn{2}{c}{998.9  }&
 \multicolumn{2}{c}{9908.6  }& \multicolumn{2}{c}{24440.5 }\\
Meta fixed  &  73.04&  36.52 &   82.63  & 8.26&      91.62 & 0.92 &   97.17  & 0.10 & 98.47 & 0.04\\
Meta random& 73.04  &  36.52 &    82.63  &  8.26 &    91.62&  0.92&    97.17 &  0.10 &   98.47 &  0.04\\
$\sum_s W_s$  &   38.53 &  19.27  &  51.37  &  5.14 &    69.54 &  0.70  &    85.59 &  0.09  &  90.61  & 0.04 \\
$-2 \sum \ln(p)$  &   41.76   &  20.88  &   54.28  &  5.43&    71.62  &  0.72 &   86.44  & 0.09 &  91.09      &
0.04\\[5pt]
 \multicolumn{11}{c}{$S=5, n_1=1000, n_s=250, s=2, \ldots, 5,  M=1$ true disease SNP }    \\
 Comb list &  22.00 & 1.10 &    33.52   &0.34 &    56.34 & 0.06 &   80.17  &  0.01 & 88.71  & 0.004  \\
Ave $T^c$  & \multicolumn{2}{c}{ 20.0}&     \multicolumn{2}{c}{100.0  }&  \multicolumn{2}{c}{999.1    }&
 \multicolumn{2}{c}{9919.8   }& \multicolumn{2}{c}{24503.9 }\\
Meta fixed  &   74.85  & 3.75 &   82.83  & 0.83 &  91.33 &  0.09 & 97.01 & 0.01 & 98.44  &  0.004\\
Meta random&  72.35 & 3.62 &   81.08& 0.81 &  90.35 &0.09 &    96.54  &0.01 & 98.05    & 0.004 \\
$\sum_s W_s$  &  54.34 & 2.73&   65.03   &0.65 &  79.49 &0.08 &  90.94  &0.01  &  94.46  & 0.004  \\
$-2 \sum \ln(p)$  &    55.72 &  2.79&   66.06 & 0.66 &   80.02 &  0.08 &   91.20  &0.01 & 94.66  &   0.004\\
 \hline
 \end{tabular*}
 \end{table*}

 \begin{table*}[t]
 \caption{Detection Probability (DP) and Proportion Positive (PP)
 for five methods for combining data from $S$
 studies,  with $n_s$  cases and $n_s$ controls   for the    random effects model  for $\beta \sim
 N(\log(1.3), 0.05^2),$ with $N=500{,}000$ SNPs, and random allele frequency
 $\eta$}\label{tab2}
 \begin{tabular*}{\tablewidth}{@{\extracolsep{4in minus 4in}}ld{2.2}d{2.2}d{2.2}ccccccd{1.3}@{}}
\hline
 \textbf{Method} &     \multicolumn{2}{c}{$\bolds{T=20}$}  &  \multicolumn{2}{c}{$\bolds{T=100}$}   & \multicolumn{2}{c}{$\bolds{T=1000}$}
  & \multicolumn{2}{c}{$\bolds{T=10{,}000}$ }  &
  \multicolumn{2}{c@{}}{$\bolds{T=25{,}000}$}\\
\ccline{2-3,4-5,6-7,8-9,10-11}
     &  \multicolumn{1}{c}{$\bolds{\mathit{DP}}$} & \multicolumn{1}{c}{$\bolds{\mathit{PP}}$} & \multicolumn{1}{c}{$\bolds{\mathit{DP}}$} & \multicolumn{1}{c}{$\bolds{\mathit{PP}}$}
  &      \multicolumn{1}{c}{$\bolds{\mathit{DP}}$} & \multicolumn{1}{c}{$\bolds{\mathit{PP}}$}  & \multicolumn{1}{c}{$\bolds{\mathit{DP}}$} & \multicolumn{1}{c}{$\bolds{\mathit{PP}}$} & \multicolumn{1}{c}{$\bolds{\mathit{DP}}$} & \multicolumn{1}{c}{$\bolds{\mathit{PP}}$}  \\
\hline
\multicolumn{11}{c@{}}{$S=5, n_s=400,    M=1$ true disease SNP }    \\
 Comb list  &  12.50  &  0.63 &  23.40&  0.23 &     48.30 &  0.05   &  77.60 &  0.01  & 88.50 & 0.004\\
Ave $T^c$  &\multicolumn{2}{c}{   20.0    }&  \multicolumn{2}{c}{ 100.0     }& \multicolumn{2}{c}{   999.0     }&\multicolumn{2}{c}{  9919.3 }&\multicolumn{2}{c}{   24503.5  }\\
Meta fixed &    73.80  & 3.69 &   82.30  &0.82 &   91.40   &0.09  &   97.50   & 0.01 &   98.60  & 0.004\\
Meta random    &   73.80   &3.69  &  82.50   &0.83&   91.40  & 0.09 & 97.50   & 0.01 &  98.60  & 0.004\\
$\sum_s W_s$  & 55.70   & 2.79 &  67.10  & 0.67  &  80.60  & 0.08 & 92.00 & 0.01 & 95.10 & 0.004\\
 $-2 \sum_s \ln(p_s)$ &  58.20  &2.91 &   68.80  &  0.69 &  81.80  &  0.08 &   92.40  &  0.01 &  95.30
 &0.004\\[5pt]
\multicolumn{11}{c@{}}{$S=5,  n_s=400,   M=10$ true disease SNPs }    \\
 Comb list  &   11.61  &   5.86 &   22.50 &  2.26  &    47.25 &  0.47 &   76.83 &0.08 &  86.36 &   0.04\\
 Ave $T^c$  &\multicolumn{2}{c}{   19.9     }&  \multicolumn{2}{c}{  99.7     }& \multicolumn{2}{c}{   997.5  }
 &\multicolumn{2}{c}{  9913.8 }&\multicolumn{2}{c}{ 24494.3  }\\
Meta fixed &    71.99  &  36.00 & 81.51  &  8.15& 90.81  & 0.91  & 97.04   &0.10 & 98.45  &  0.04\\
Meta random    &  71.96  & 35.98 &   81.50 & 8.15&  90.74  &  0.91 &  97.04  & 0.10 &  98.45  &  0.04\\
$\sum_s W_s$  & 54.85  &27.43  &  66.06  & 6.61  &  79.84  & 0.80 &  91.14   & 0.09 &   94.55  &  0.04\\
 $-2 \sum_s \ln(p_s)$ &   57.39  &  28.70 &   67.91  & 6.79 &  81.15   & 0.81&  91.73   &  0.09&  94.88  &
 0.04\\[5pt]
\multicolumn{11}{c@{}}{$S=10, n_s=200,   M=1$ true disease SNP }    \\
 Comb list  &   2.00  &    0.10 & 4.70    &  0.05 &    18.90   &  0.02  &  54.80    &  0.06 &   70.40     &  0.003\\
Ave $T^c$  &\multicolumn{2}{c}{   20.0    }&  \multicolumn{2}{c}{ 100.0     }& \multicolumn{2}{c}{   999.0     }
&\multicolumn{2}{c}{ 9910.5 }&\multicolumn{2}{c}{   24444.3   }\\
Meta fixed &   74.10   &  3.71  &  82.30   &   0.82 & 92.00  & 0.09 &  97.50   &  0.01  &  98.90   & 0.004 \\
Meta random    &  74.10  &  3.71  &  82.30  &  0.82  &  92.00   &  0.09  &  97.50 &  0.01&   98.90  &  0.004\\
$\sum_s W_s$  &  42.00 &  2.10 &   52.80  &  0.53 &   69.20   & 0.07 &  85.50    &  0.01 &  90.30   & 0.004\\
 $-2 \sum_s \ln(p_s)$ &  44.70   &  2.24 &   55.00  & 0.55 &  71.10 &  0.07 &   86.40 &  0.01 &  91.30   &
 0.004\\[5pt]
\multicolumn{11}{c@{}}{$S=10, n_s=200,  M=10$ true disease SNPs }    \\
 Comb list  &   2.03 &  1.02   &   5.75 & 0.58   &  20.47  &  0.21  &  54.12  &   0.05  & 70.32  &   0.03\\
Ave $T^c$  &\multicolumn{2}{c}{   20.0    }&  \multicolumn{2}{c}{ 100.0     }& \multicolumn{2}{c}{   998.8    }
&\multicolumn{2}{c}{  9907.8}&\multicolumn{2}{c}{  24440.4  }\\
Meta fixed &   72.24   &  36.12&  81.74  &  8.17 & 91.57  &   0.92 &   96.92  &  0.10 & 98.50  &  0.04\\
Meta random    &  72.22   &  36.11 &  81.73  & 8.17 &  91.55 &  0.92 & 96.89   & 0.10  & 98.50  &  0.04\\
$\sum_s W_s$  &  41.81   &  20.91 &  54.18   &  5.42  &   70.95   &  0.71 &   85.74  & 0.09  & 90.79  & 0.04\\
 $-2 \sum_s \ln(p_s)$ &  44.71  & 22.36 &  56.42  & 5.64 &  72.43 &  0.72 & 86.48   &  0.09 & 91.17 &
 0.04\\[5pt]
 \multicolumn{11}{c@{}}{$S=5, n_1=1000, n_s=250, s=2, \ldots, 5,  M=1$ true disease SNP }    \\
 Comb list  &    25.70  & 1.29  &   38.20 &0.38  & 57.90  &  0.06 &     81.10 & 0.01 &  88.90   &    0.004\\
Ave $T^c$  &\multicolumn{2}{c}{   20.0    }&  \multicolumn{2}{c}{ 100.0     }& \multicolumn{2}{c}{  999.1    }
&\multicolumn{2}{c}{  9919.7 }&\multicolumn{2}{c}{  24503.8}\\
Meta fixed &     74.70 &  3.74  &    83.10 & 0.83  &    91.90& 0.09  &    97.20 &0.01  &  98.60 & 0.004\\
Meta random    &   72.30& 3.62 &     81.80 &0.82  &     91.10  &  0.09   &     96.50  &0.01 &  98.20 & 0.004\\
$\sum_s W_s$  &   55.80  & 2.79&   66.80  & 0.67  &    81.10 & 0.08&      91.40 &0.01 &   94.70  &0.004\\
 $-2 \sum_s \ln(p_s)$ &56.80  &  2.84 &   67.90  &0.68 &    82.30  &0.08  &      91.80   & 0.01 & 94.80  &0.004\\
 \hline
 \end{tabular*}\vspace*{4pt}
\end{table*}

\begin{table*}[t]
 \caption{Detection Probability (DP) and Proportion Positive (PP)
 for five methods for combining data from $S$ studies, with $n_s$
  cases and $n_s$ controls  for the    random effects model  for $\beta \sim
    N(\log(1.3), 0.5^2),$ with $N=500,000$ SNPs, and random allele
    frequency~$\eta$}\label{tab3}
 \begin{tabular*}{\tablewidth}{@{\extracolsep{4in minus 4in}}ld{2.2}d{2.2}cd{2.2}cccccd{1.3}@{}}
\hline
 \textbf{Method} &     \multicolumn{2}{c}{$\bolds{T=20}$}  &  \multicolumn{2}{c}{$\bolds{T=100}$}   & \multicolumn{2}{c}{$\bolds{T=1000}$}
  & \multicolumn{2}{c}{$\bolds{T=10{,}000}$ }  &
  \multicolumn{2}{c@{}}{$\bolds{T=25{,}000}$}\\
  \ccline{2-3,4-5,6-7,8-9,10-11}
 &  \multicolumn{1}{c}{$\bolds{\mathit{DP}}$} & \multicolumn{1}{c}{$\bolds{\mathit{PP}}$} & \multicolumn{1}{c}{$\bolds{\mathit{DP}}$} & \multicolumn{1}{c}{$\bolds{\mathit{PP}}$}
  &      \multicolumn{1}{c}{$\bolds{\mathit{DP}}$} & \multicolumn{1}{c}{$\bolds{\mathit{PP}}$}  & \multicolumn{1}{c}{$\bolds{\mathit{DP}}$} & \multicolumn{1}{c}{$\bolds{\mathit{PP}}$} & \multicolumn{1}{c}{$\bolds{\mathit{DP}}$} & \multicolumn{1}{c}{$\bolds{\mathit{PP}}$}  \\
\hline
  \multicolumn{11}{c@{}}{$S=5, n_s=400,    M=1$ true disease SNP }    \\
    Comb list &      86.10 &4.54 &   89.50  &0.91   &  94.40& 0.09&     98.10 & 0.01 & 98.30  &0.004\\
Ave $T^c$    &   \multicolumn{2}{c}{  19.1}&  \multicolumn{2}{c}{     99.0 }& \multicolumn{2}{c}{  997.9  }&
\multicolumn{2}{c}{  9918.8 }& \multicolumn{2}{c}{ 24503.6  }\\
Meta fixed  &      57.90 &    2.90 &   62.50 &0.63 &       70.20 & 0.07  &  77.70  & 0.01  & 81.50 &0.003\\
 Meta random&  55.90& 2.80 &    60.20  & 0.60 &    66.20  &0.07 &     75.80  & 0.01 &   80.00 &0.003 \\
 $\sum_s W_s$  &   93.20& 4.66 &     94.80  & 0.95 &    97.00  & 0.10 &   98.20  & 0.01&  98.70  & 0.004\\
   $-2 \sum_s \ln(p_s)$   &    93.40& 4.67 &    94.70 & 0.95   &   97.20 & 0.10 &     98.20  & 0.01 &   98.80 &
   0.004\\[5pt]
\multicolumn{11}{c@{}}{$S=5, n_s=400,   M=10$ true disease SNPs }    \\
 Comb list &      56.74 &  36.62&    89.29 & 10.12 &   94.53 &  0.96 &  97.46&  0.10&   98.21 &  0.040\\
Ave $T^c$ &\multicolumn{2}{c}{  16.6 }& \multicolumn{2}{c}{   89.9 }& \multicolumn{2}{c}{ 985.8  }&
 \multicolumn{2}{c}{ 9903.5 }& \multicolumn{2}{c}{24486.0}\\
Meta fixed  &   58.35 &29.18 &   63.70  & 6.37  &  70.85 & 0.71 &   78.73 & 0.08 &  81.94  & 0.033\\
Meta random&   55.17 & 27.59 &   60.33 & 6.03 &    67.74 &0.68&   75.93& 0.08 & 79.70 & 0.032\\
$\sum_s W_s$  &      92.46 & 46.23&   94.79 &9.48 &  96.77&   0.97 &  98.32  & 0.10 &  98.92  &0.040\\
 $-2 \sum_s \ln(p_s)$  & 92.36  & 46.18 &   94.68 &9.47 &    96.73 &0.97 &     98.36 &0.10  & 98.87
 &0.040\\[5pt]
\multicolumn{11}{c@{}}{$S=10, n_s=200,  M=1$ true disease SNP }    \\
 Comb list  &  84.70  & 4.52 &  89.30  & 0.91  &   94.70&    0.10 & 97.90 & 0.01 &  98.60&   0.004\\
Ave $T^c$  &\multicolumn{2}{c}{  19.0  }&  \multicolumn{2}{c}{ 98.7  }& \multicolumn{2}{c}{ 997.1   }&
\multicolumn{2}{c}{9907.8} &\multicolumn{2}{c}{24441.4}\\
Meta fixed &  67.30 & 3.37 &    72.00 &   0.72 &   78.80 & 0.08 &    85.60&    0.01 & 88.30&  0.004\\
Meta random    & 60.60  & 3.03   & 66.30   & 0.66 &  73.00 &  0.07 &   82.00&  0.01 &  85.20  &  0.003\\
$\sum_s W_s$  &   96.20 &  4.81& 96.90 &   0.97 &     98.50 & 0.10 &     99.40&  0.01 &  99.70 &  0.004\\
 $-2 \sum_s \ln(p_s)$ &   96.30 & 4.82 &   97.10 &  0.97 &  98.40 & 0.10 &    99.30  &  0.01 & 99.70 &  0.004
 \\[5pt]
\multicolumn{11}{c@{}}{$S=10, n_s=200,   M=10$ true disease SNPs }    \\
  Comb list &45.52   &26.12 &  86.79  &10.12 &    94.60  &  0.10 &  98.00  &0.10  &  99.00  &0.041\\
Ave $T^c$ &\multicolumn{2}{c}{ 17.9 }&  \multicolumn{2}{c}{   87.7}& \multicolumn{2}{c}{  979.9}&
\multicolumn{2}{c}{  9884.0 }& \multicolumn{2}{c}{ 24415.5}\\
Meta fixed  &65.10  &32.55 &   70.74  &  7.07 &   77.86  &0.78&     84.85 & 0.09&   88.03  &0.035\\
Meta random&  59.61 & 29.81&    65.21  & 6.52&    73.13  &0.73 &    80.93   &0.08  & 84.29 & 0.034\\
$\sum_s W_s$  &95.45 & 47.73 &  96.97 &9.70  &   98.35  & 0.98&   99.26  &   0.10& 99.59    &0.040\\
 $-2 \sum_s \ln(p_s)$ &95.24   &47.62 &   96.82   &9.68 &   98.34 &0.98 &   99.21  &  0.10  & 99.55
 &0.040\\[5pt]
\multicolumn{11}{c@{}}{$S=5, n_1=1000, n_s=250, s=2, \ldots, 5,  M=1$ true disease SNP }    \\
 Comb list  &    83.80  &  4.37 &     87.60   &  0.88   &     92.80  & 0.09  &       96.70  &  0.01 &   97.90  &   0.004\\
Ave $T^c$  &\multicolumn{2}{c}{  19.3     }&  \multicolumn{2}{c}{ 99.2      }& \multicolumn{2}{c}{   998.1    }
&\multicolumn{2}{c}{   9918.5}&\multicolumn{2}{c}{  24501.9}\\
Meta fixed &    60.10   &   3.01&   64.00  &  0.64  &      69.00  &  0.07 &      76.40    & 0.01 &   81.20 &   0.003\\
Meta random    &   50.80   &   2.54 &   55.70    &  0.56 &     62.20  &  0.06  &      71.40 & 0.01 &   75.70   &   0.003 \\
$\sum_s W_s$  &   90.30   &  4.52  &      92.80  &  0.92  &     95.50   &  0.10  &     97.80  & 0.01&  98.90  & 0.004 \\
 $-2 \sum_s \ln(p_s)$ & 90.20  &  4.51   &    92.30   & 0.92 &      95.60   &  0.10  &     97.60  & 0.01 &   98.80    &     0.004 \\
         \hline
 \end{tabular*}\vspace*{3pt}
\end{table*}

\subsection{Simulation Results for Detection Probability}\label{sec5.3}

We evaluated the $\mathit{DP}$ for $T = 20, 100, 1000, 10{,}000$ and $25{,}000,$
which, when divided by $N,$ corresponds to respective selection
fractions 0.00004, 0.0001,\break 0.0005, 0.02
and 0.05. We studied $M=1$ and $M=10$ disease SNPs, and let $S=5$ with
$n_s=400$ cases and controls and $S=10$ with $n_s=200$ cases and controls
for both the fixed and the random effects models for $\beta$, and we
focused on
$\beta=\log(1.3)$. To assess the impact of varying study sizes, with
$S=5$, we let
$n_1=1000$ and $n_s=250, s=2,\ldots,5$.

For the fixed effects model (Table \ref{tab1}), the two meta-analytic approaches
had the highest $\mathit{DP}$ for all study designs, followed by Fisher's
combination of $p$-values and then the sum of the Wald statistics.
The ``combined list approach'' had the lowest
$\mathit{DP}$ of all approaches. For example, for $T=20$, $\mathit{DP}$ for the list was
only 7.2\% for five studies with $n_s=400$ cases and $400$ controls
each, and a single true disease-associated SNP, $M=1$, while {DP} was
53.9\% and 58.4\% for the sum of Wald tests and the Fisher $p$-value
combination respectively, and 74.2\% for both meta-analytic approaches.
In the same setting, for $T=25{,}000$, {DP} for the combined list approach
was 85.3\%, while it was 94\% or higher for all other approaches (Table
\ref{tab1}). For $S=10$ and $n_s=200$, the combined list approach had even
smaller $\mathit{DP}$ values, because each of the component studies had a very
small $\mathit{DP}$. Similar patterns were observed for $M=10$. The number of
disease-associated SNPs, $M$, did not strongly impact {DP} for any of the
methods under the fixed effects model.
For $S=5$ and varying study sizes, $n_1=1000$ and $n_s=250, s=2,\ldots
,5$, for $M=1$, the performance of the combined list approach was
slightly better, with $\mathit{DP}=22.0$\% for $T=20$, because study $s=1$ had a
larger size and higher $\mathit{DP}$.

The proportions positive ($\mathit{PP}$) were largest for small $T$ and larger
$M$. As $T$ increased, $\mathit{DP}$ increased but $\mathit{PP}$ declined (Table~\ref{tab1}). If
the purpose of the study is to serve as an initial screen designed to
capture disease SNPs but tolerate a large number of false positive
results (i.e., very small $\mathit{PP}$), $T=25{,}000$ might be of interest. If
the purpose is to select a small number of promising SNPs for further
study, data for $T=20$ commend the meta-analytic approaches.
For the settings we studied, the Liptak--Stouffer combination of
$p$-values had a lower $\mathit{DP}$ than Fisher's combination of $p$-values. For
example, for $S=10$ and $n_s=400$, with $M=1$ true disease-associated
SNP, the values of
$\mathit{DP}$ were $55.5\%, 64.8\%,\break 76.2\%, 86.9\%$ and $91.2\%$ for the
Liptak--Stouffer combination for
$T=20, 100, 1000, 10{,}000 $ and $25{,}000$,\break while the corresponding $\mathit{DP}$
values of the Fisher combination were
$ 58.4\%, 66.9\%, 80.2\%, 90.8\%$ and\break $94.5\%$. Therefore, we did not
tabulate results for the Liptak--Stouffer combination of $p$-values.

For the random effects model (Table \ref{tab2}) with a relatively small between
study standard deviation, $\tau=0.05$, and with $\beta=\log(1.3) $
for the disease-associated SNPs, the {DP} results were very similar to
the fixed effects model. Again, the meta-analytic approaches had better
$\mathit{DP}$ than the combined list, sum of Wald tests, or Fisher $p$-value
combinations. However, for the random effects model with a very large
standard deviation, $\tau=0.5$ (Table~\ref{tab3}), Fisher's combination of
$p$-values and the sum of the Wald statistics had much better {DP} than the
meta-analytic approaches, as the large variation among the $\hat\beta
^s$ for the disease-associated SNPs caused some of them to be negative,
reducing the meta-analytic estimate of the overall effect (Table~\ref{tab3}).
For $\tau=0.5$ the combined list approach also had higher $\mathit{DP}$ than
the two meta-analytic approaches. Even for $T=25{,}000$, for $S=5$
studies with $400$ cases and $400$ controls each, and a single true
disease-associated SNP, $M=1$, {DP} was 81.5\% and 80.0\% for the fixed
and random effects meta-analytic approaches, compared to 98.3\%, 98.7\%
and 98.8\% for the combined list, the sum of Wald statistics and
Fisher's combination of $p$-values (Table~\ref{tab3}). For $T=20$, $\mathit{DP}$ for the
combined list approach was considerably lower when the number of
disease-associated SNPs was $M=10$, because in each study the 10
disease SNPs compete against each other for only
$T/S=4$ top positions. This competition is less pronounced in
Tables~\ref{tab1}
and~\ref{tab2} because the magnitude of log-odds ratios for disease-associated
SNPs does not reach the large values that sometimes occur in
simulations in Table~\ref{tab3} with $\tau=0.5$.
Similar to the fixed effects setting, the Liptak--Stouffer combination
of $p$-values had a lower $\mathit{DP}$ than Fisher's combination of $p$-values and
the sum of Wald tests for the random effects models with $\tau=0.05$
and $\tau=0.5$ and, therefore, we did not tabulate these results.

For fixed effects models (Table \ref{tab1}), studies with $S=5$ and $n_s=400$
resulted in higher $\mathit{DP}$ than studies with the same total number of
subjects but $S=10$ and $n_s=200$ for the combined list, the sum of
Wald statistics and Fisher's combination of $p$-values, for both $M=1$
and $M=10$ disease SNPs; no such difference was seen for the
meta-analytic approaches. Under the random effects model with $\tau
=0.05$ (Table~\ref{tab2}), $\mathit{DP}$ was higher for the combined list, sum of Wald
statistics and Fisher's combination of $p$-values for $S=5$ with
$n_s=400$ than for $S=10$ with $n_s=200$. In this case the
meta-analytic procedures had comparable or slightly higher {DP} for
$S=10, n_s=200$. Under the random effects model with $\tau=0.5$ (Table
\ref{tab3}), all procedures except the combined lists had higher {DP} with $S=10, n_s=200$.

\begin{figure*}[t]

\includegraphics{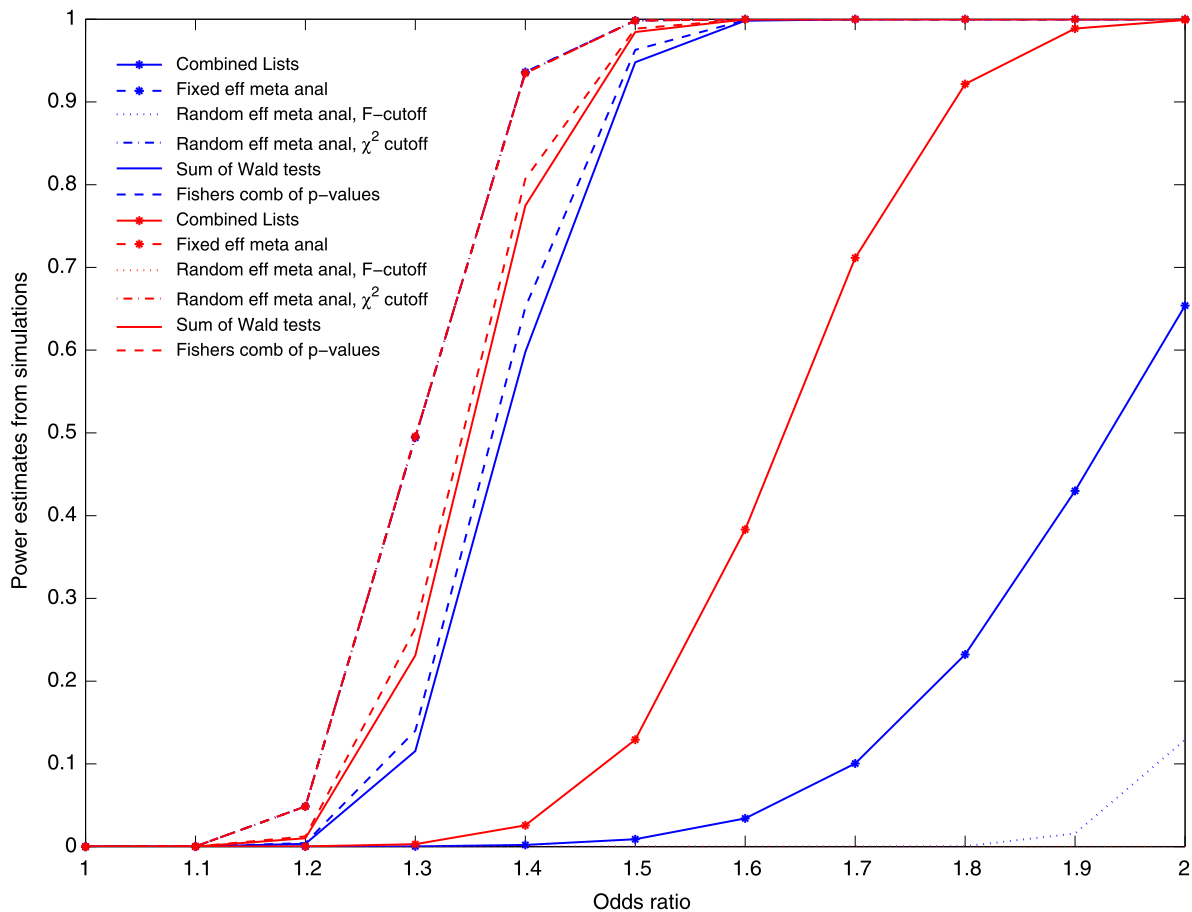}

  \caption{Power of various approaches for combing data from $S=10$ GWAS
studies with $n_s =200$ cases and $n_s=200$ controls each (blue lines) or
$S=5$ studies with $n_s =400$ cases and $n_s =400$ controls each (red lines) under the
fixed effects model for disease-associated SNPs, with $\eta=0.2673$.}\label{fig1}
\end{figure*}

\begin{figure*}[t]

\includegraphics{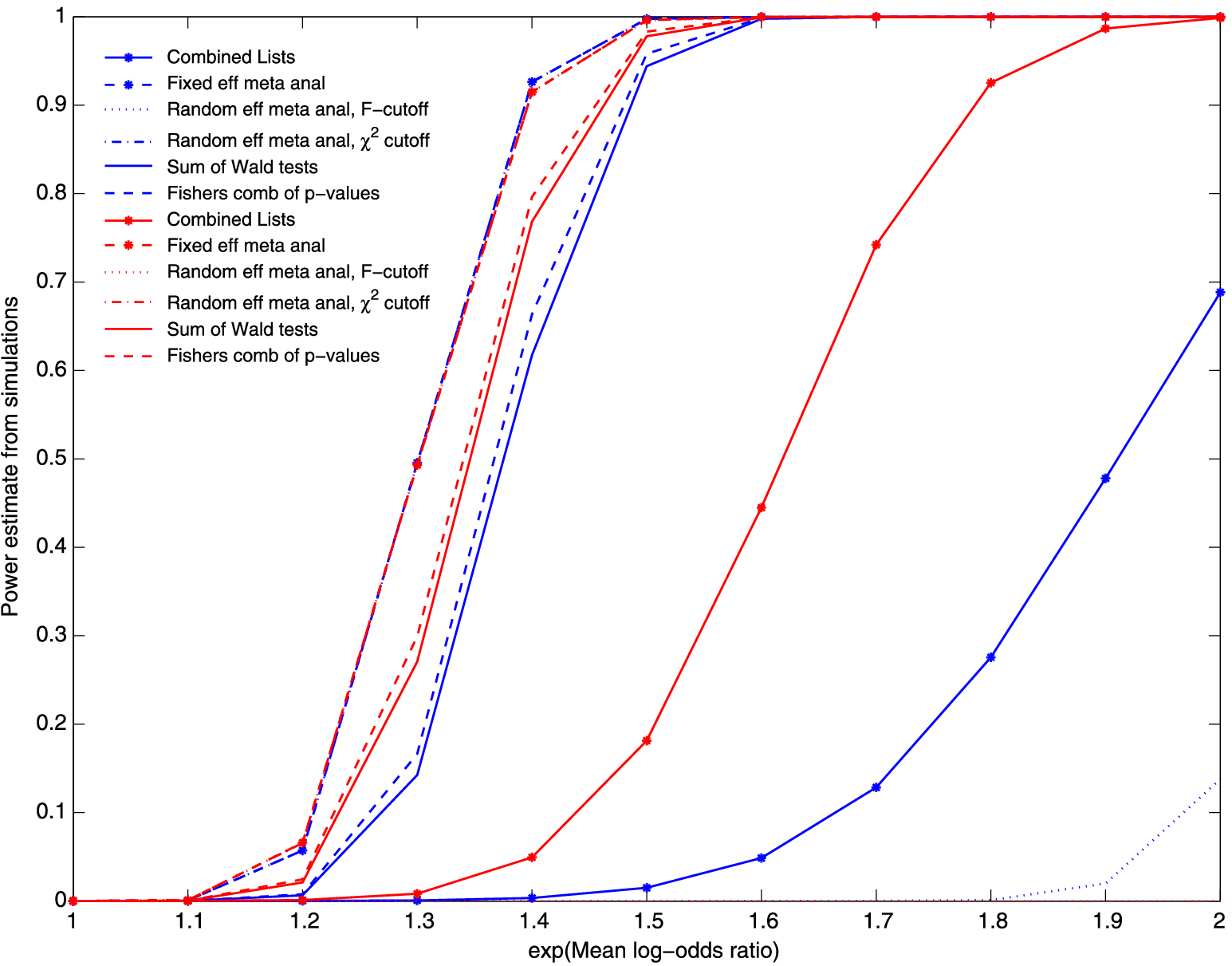}

  \caption{Power of various approaches for combing data from $S=10$ GWAS
studies with $n_s =200$ cases and $n_s =200$ controls each (blue lines) or
$S=5$
studies with $n_s =400$ cases and $n_s =400$ controls each (red lines) under the
random effects model for disease-associated SNPs, $\beta^{s}\sim N(\beta,
0.05^{2})$, with $\eta=0.2673$.}\label{fig2}
\end{figure*}

\begin{figure*}[t]

\includegraphics{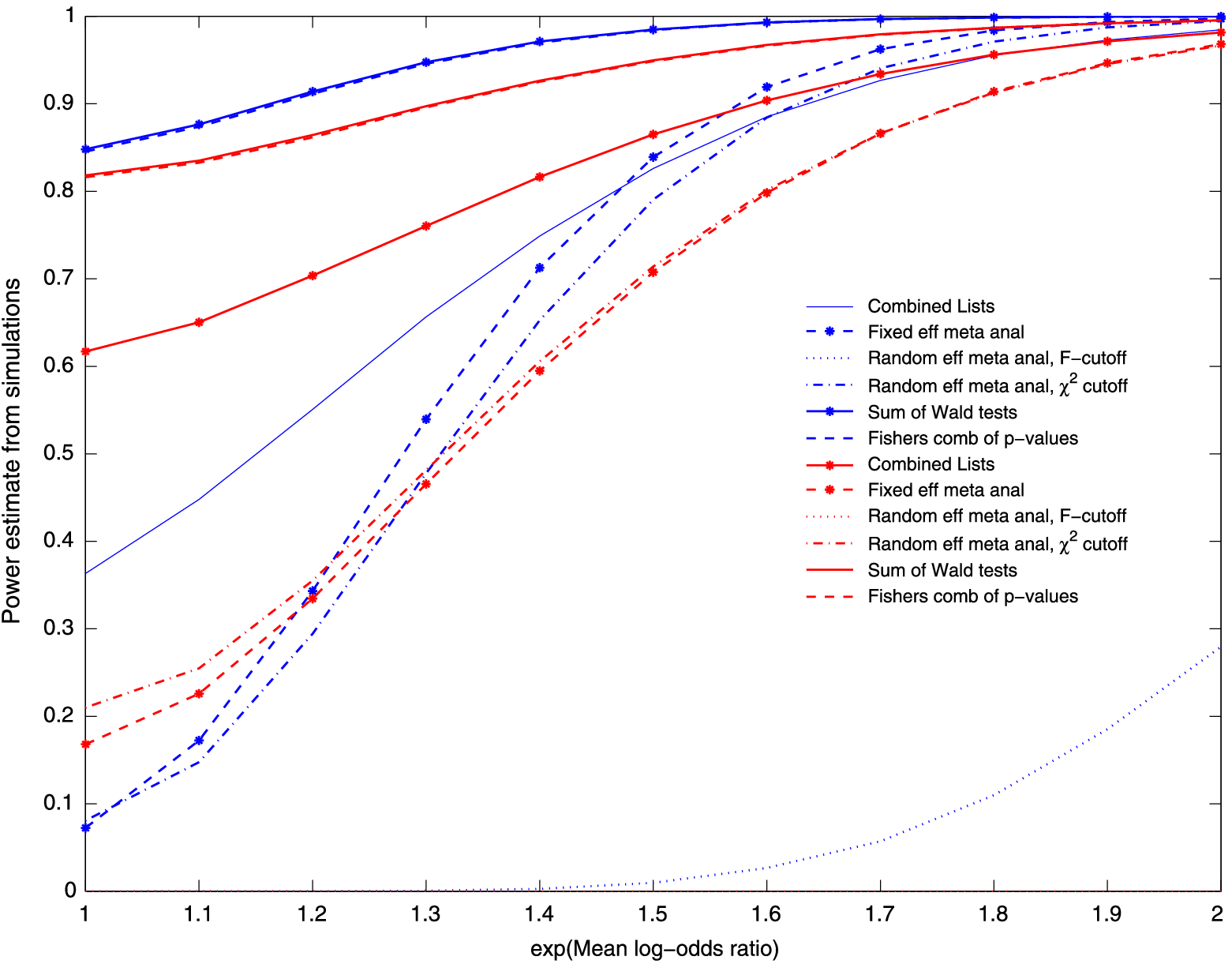}

  \caption{Power of various approaches for combing data from $S=10$ GWAS
studies with $n_{s}=200$ cases and $n_{s}=200$ controls each (blue
lines) or $S=5$ studies with $n_{s} =400$ cases and $n_{s} =400$ controls each (red
lines) under the random effects model for disease-associated SNPs,
$\beta^{s}$$\sim$$N(\beta, 0.5^{2})$, with $\eta=0.2673$.}\label{fig3}\vspace*{5pt}
\end{figure*}

\subsection{Simulation Results for Power}\label{sec5.4}

Power estimates based on $\mathit{NSIM}=100{,}000$ simulations are plotted against
odds ratios (Figure~\ref{fig1}) for $S=5 $ with $n_s=400$ and for $S=10$ with
$n_s=200$ under the fixed effects model. The odds ratio was assumed to
be the same in all $S$ studies. For all combinations of $S$ and $n_s$,
the fixed effects meta-analytic approach had the largest power for all
odds-ratios. It gave the exact same results as the random effects
meta-analytic approach with the critical region defined by the $\chi
^2_{1,1-\alpha}$ quantile, leading to indistinguishable lines in
Figure~\ref{fig1}. Using the $F_{1,S-1,1-\alpha}$ cutoff value for the random
effects meta-analytic approach resulted in extremely low power.
Additionally, for the meta-analytic approaches, $S=5$ with $n_s=400$
resulted in the exact same power as $S=10$ with $n_s=200$, as the total
sample size was the same. The sum of Wald-test statistics and Fisher's
$p$-value combination gave very similar results with 80\% power for odds
ratios near 1.4 compared to 93\% power for the meta-analytic
approaches. The power of the combined list approach was noticeably
lower, and reached 80\% only for an odds ratio~$=$ 1.75. These empirical
power estimates agreed well with the analytic power calculations (data
not shown).

For the random effects model for the disease-asso\-ciated SNPs, $\beta^s
\sim N(\beta,\tau^2)$,
with a small random effects standard deviation, $\tau=0.05$, the
estimated power of these
procedures was very similar to their power under the fixed effects
model (Figure~\ref{fig2}). If
the random effects standard deviation was $\tau=0.5$, there was enough
heterogeneity in
association effects across studies that the log odds were positive in
some studies and
negative in others, leading to a reduction in the meta-analytic summary
estimate of association,
and to substantial loss in power compared to all other procedures
(Figure~\ref{fig3}). For example, for
$S=5$ with $n_s=400$ (Figure~\ref{fig3}), an expected log-odds ratio of $\log
(1.6)$ was required to
attain 80\% power for the meta-analytic approach. On the other hand,
the sum of Wald tests
or Fishers combination are invariant to sign changes of the effects,
and had very high power.
For example, even for mean log-odds ratio $\beta=0$, the power of
those two procedures was near
80\% for $S=10$ with $n_s=200$ and $S=5$ with $n_s=400$. The combined
list procedure also had
much higher power than the meta-analytic approaches, for example, 82\%
for a mean log-odds ratio
of $\log(1.4)$ for $S=5$ with $n_s=400$. Again, for the fixed effects
meta-analysis and the
random effects meta-analysis with the critical region defined by the
$\chi^2_{1,1-\alpha}$
quantile, the lines completely overlap and are indistinguishable in
Figures~\ref{fig2} and~\ref{fig3}.

The power of the Liptak--Stouffer combination of $p$-values for all
settings studied for the figures was very close to the power of the
Fisher statistic and therefore is not presented. For example, for the
fixed effects model presented in Figure \ref{fig1}, for an $\mathit{OR}=1.5$, with $200$
cases and $200$ controls for $10$ studies, the power of the Fisher
combination was $0.9581$ and for the Liptak--Stouffer combination was
$0.9535$. For $400$ cases and $400$ controls and $5$ studies, the power
for an $\mathit{OR}=1.4$ was $0.8057$ for Fisher's and $0.8167$ for the
Liptak--Stouffer combination of $p$-values.

Fewer studies with larger sample size ($S=5,n_s=400$) resulted in
higher power than more studies with the same total number of subjects
($S=10$ and $n_s=200$) for all procedures (with the exception of the
meta-analytic approaches, for which the power was the same) under the
fixed effects model and under the random effects model with $\tau
=0.05$ (Figures~\ref{fig1} and~\ref{fig2}). When $\tau=0.5$, however, the power of all
approaches but the combined list was larger for $S=10$ studies with
$n_s=200$ (Figure~\ref{fig3}).

\section{Discussion}\label{sec6}

As is evident from the literature on detection probability (Gail et
al., \citeyear{2008agail1}, \citeyear{2008bgail2}) and power
calculations (Skol et al., \citeyear{2006skiol2}, \citeyear{2007skol}), large sample sizes are needed
to have a good chance to
discover disease-associated SNPs with odds ratios commonly found in GWA
studies. Because in
many settings the available studies are too small, there is a need to
combine information from
several studies. Our results indicate that the fixed effects
meta-analysis has higher {DP} than
other methods. Only when there is severe heterogeneity in association
effects across studies
such that the log odds is positive in some studies and negative in
others can methods such as
sum of Wald tests or Fishers combination of $p$-values have larger {DP}
than the fixed effects and
random effects meta-analytic approaches.

Loughin (\citeyear{2004lough}) found, in an extensive simulation study of the power of
various quantile combinations methods for $p$-values,
that
Fisher's method had very good power compared to other transformation
functions (including normal and logistic) when
a minority of the tests provided most of the evidence against the null
hypothesis.
When signal was distributed equally over all $p$-values, the normal
transformation proved to be somewhat more powerful than Fisher's
approach. We therefore also assessed the performance of the
Liptak--Stouffer combination of $p$-values. In our simulation studies,
under both the fixed effects and
the random effects model for the disease associated SNPs, Fisher's
combination of $p$-values had higher $\mathit{DP}$ than the Liptak--Stouffer
combination of $p$-values, but had very similar power.

Although differences in LD patterns across populations can result in
associations in opposite directions, as illustrated by CDKN1AS31R, in
the supplement to Zeggini et al. (\citeyear{2008zegg}), in most circumstances the
heterogeneity will not be sufficient to render the meta-analytic
approaches less powerful than other approaches.
The method of combining lists of promising SNPs from each of the
component studies has the lowest {DP} in most circumstances, and
especially when there are many small studies of comparable size. Our
results for power give a similar ranking of procedures to combine
information as for {DP}, despite the fact that these two criteria are far
from equivalent (Gail et al., \citeyear{2008bgail2}).

We used the critical values from a one degree-of-freedom chi-square
distribution in power calculations for the random effects meta-analytic
procedure discussed by DerSimonian and Laird (\citeyear{1986der}). Under the strong
null hypothesis that the log odds is strictly zero, we conducted
simulations and verified that such critical values yielded proper size
in simulations for $\alpha=0.1$ and $\alpha=0.01$. It is not certain
that the size is nominal for $\alpha=10^{-7}$, however, and therefore
the power from the random effects meta-analytic approach may not be
strictly comparable to that of the fixed effects meta-analysis. If in
fact null SNPs satisfy only a weak null hypothesis, namely, that their
log odds have mean zero but vary about this mean, then a critical value
based on an $F$ distribution might be more appropriate (Follmann and
Proschan, \citeyear{1999foll}). Using such a critical value reduces power to almost
zero, however, as shown in Figures~\ref{fig1}, \ref{fig2} and~\ref{fig3}. In Section~\ref{sec2} we argue
that a strong null hypothesis is plausible.

We assumed that the same platform was used to analyze the samples in
each study and thus that data were available on the same set of SNPs in
each study. Zeggini et al. (\citeyear{2008zegg}) used two algorithms that employed
Hapmap data to impute missing SNPs in some studies. We also assumed
that adequate quality control procedures had been followed in all the
studies and that there was proper control for population
stratification. Otherwise, the assumption of a strong null hypothesis
for nondisease-associated SNPs would not hold.

\appendix
\section*{Appendix}\label{app}

\subsection*{Variance Computation for Model (\protect\ref{model1})}
\renewcommand{\theequation}{\arabic{equation}}
\setcounter{equation}{14}

For ease of exposition we omit the SNP specific subscript, and denote
(\ref{model1}) by
$p^s_{x}=1-q^s_{x}=P(Y=1|X=x; \mu^*_s,\beta),$ for $s=1,\ldots, S$.
The maximum likelihood estimate $\hat\beta$ is found by solving the
score equations corresponding to the likelihood (\ref{likelihood1}),
\begin{eqnarray}
 \partial/ \partial\mu_s \log L &=& \sum_j (Y_{sj} - p^s_{xj})=0,
\\
\eqntext{\quad s=1, \ldots, S,}
\\
 \partial/ \partial\beta\log L &=& \sum_s \sum_j x_{sj}(Y_{sj} -
p^s_{xj})=0,
\end{eqnarray}
where the index $j$ refers to the $j$th subject in study $s$.
The first set of equations corresponds to the study specific intercept
parameters, and the last equation corresponds to the common log-odds
ratio parameter $\beta$. The variance $\sigma^2_S = \operatorname{var}(\hat\beta)
= (I_{22} - I_{21} I_{11}^{-1} I_{12})^{-1}$, where $I_{11}, I_{12},
I_{22}$ are submatrices of the information matrix $I$ from the
prospective likelihood:
\begin{eqnarray*}
(I_{11})_{ij} &=& E(\partial^2 / \partial\mu_i\, \partial\mu_j \log
L), \\[2pt]
(I_{21})_{j} &= &E(\partial^2 / \partial\mu_j\, \partial\beta\log
L),\\[2pt]
I_{22} & = & -E(\partial^2 / \partial^2 \beta\log L ).
\end{eqnarray*}
The expectations of the second derivatives and cross-derivatives of the
prospective log-likelihood are taken with respect to retrospective
sampling distributions $f^s_{x}=P_s(X=x|Y=1)$ and
$g^s_{x}=P_s(X=x|Y=0)$, for cases and controls respectively.

As the studies are independent, $I_{11}$ is a diagonal matrix with the
expected second derivatives of the study specific intercept parameters
on the diagonal. Thus, the information matrix reduces to
\begin{eqnarray}
\label{fisher}
  I_{22} & = & \sum_s I_{22,s}
  \\
  &=& \sum_s \sum_{x=0}^2 n_s (f^s_{s} +
g^s_{x}) x^2 p^s_x q^s_x,\nonumber
 \\
\qquad (I_{21})_{s} &= &(I_{12})
_{s} = n_s \sum_{x=0}^2 (f^s_{x} + g^s_{x}) x
p^s_x q^s_x,
\\
 (I_{11})_{ss} &=& n_s \sum_{x=0}^2 (f^s_{x} + g^s_{x}) p^s_x q^s_x.
\end{eqnarray}
The variance for $\hat\beta$ is then given by
\begin{eqnarray}
\label{var1}
\qquad \sigma^2_S &=& \operatorname{var}(\hat\beta) = (I_{22} - I_{21} I_{11}^{-1}
I_{12})^{-1}\nonumber
\\[-8pt]\\[-8pt]
&=&
\Biggl\{ \sum_{s=1}^S [I_{22,s} - I_{21,s} (I_{11,s})^{-1} I_{12,s}] \Biggr\}^{-1}.\nonumber
\end{eqnarray}

For $S=1$ (\ref{var1}) reduces to the standard case-control variance,
\begin{equation}
\label{var0}
\sigma^2_1=
( I_{22} - I_{21} (I_{11})^{-1} I_{12})^{-1}.
\end{equation}

\subsection*{Variance Computation for the Fixed Effects Meta-Analytic Approach}
Recall that $\hat\beta^F= \sum_{s=1}^S \hat\beta^s w_s,$
where $w_s = 1/ \sigma^2_s\cdot \break (\sum_{k=1}^S 1/ \sigma^2_k)^{-1}$ and,
thus, $\operatorname{var}(\hat\beta^F)=(\sum_{s=1}^S 1/ \sigma^2_s)^{-1}$.
Using (\ref{fisher}) for a single study,
\begin{displaymath}
\sigma^2_s = \operatorname{var}(\hat\beta_{s}) = (I_{22,s} - I_{21,s} I_{11,s}^{-1}
I_{12,s})^{-1},
\end{displaymath}
where $I_s$ stands for the study specific Fisher information matrix.
Therefore,
\begin{eqnarray}
\sum_{s=1}^S 1/ \sigma^2_s = \sum_{s} (I_{22,s} - I_{21,s}
I_{11,s}^{-1} I_{12,s})
\end{eqnarray}
\eject
and, thus, $\operatorname{var}(\hat\beta^F)=(\sum_{s=1}^S 1/ \sigma^2_s)^{-1}$
equals equation~(\ref{var1}).

\vspace*{6pt}
\section*{Acknowledgment}
\vspace*{3pt}
 We thank the reviewer for helpful suggestions.


\vspace*{6pt}
\begin{thebibliography}{9}
\vspace*{6pt}
\bibitem[\protect\citeauthoryear{Arm}{1955}]{1955arm}
\textsc{Armitage, P.} (1955). Tests for linear trends in proportions and frequencies.
\textit{ Biometrics } \textbf{11} 375--386.


\bibitem[\protect\citeauthoryear{der}{1986}]{1986der}
\textsc{DerSimonian, R.} and \textsc{Laird, N.} (1986). Meta-analysis in
clinical trials. \textit{Control. Clin. Trials } \textbf{7} 177--188.

\bibitem[\protect\citeauthoryear{Arm}{1999}]{1999dev} \textsc{Devlin,
B.} and \textsc{ Roeder, K.} (1999). Genomic control for association
studies. \textit{Biometrics} \textbf{55} 997--1004.

\bibitem[\protect\citeauthoryear{Arm}{2007}]{2007east} \textsc{Easton,
D. F.,} \textsc{Pooley, K. A.}, \textsc{Dunning, A. M.} et al.
(2007). Genome-wide association study identifies\break novel breast cancer
susceptibility loci. \textit{Nature} \textbf{28} 1087--1093.

\bibitem[\protect\citeauthoryear{Fish}{1932}]{1932Fisher} \textsc{Fisher,
R. A}. (1932).  \textit{Statistical Methods for Research Workers}, 4th ed. Oliver
and Boyd, London.

\bibitem[\protect\citeauthoryear{Foll}{1999}]{1999foll} \textsc
{Follmann, D. A.} and \textsc{Proschan, M. A}. (1999). Valid inference
in random effects meta-analysis. \textit{Biometrics} \textbf{55} 732--737.

\bibitem[\protect\citeauthoryear{Gail}{2008a}]{2008agail1} \textsc{Gail,
M. H.}, \textsc{Pfeiffer, R. M.}, \textsc{Wheeler, W.} and \textsc
{Pee, D}. (2008a). Probability of detecting disease-associated single
nucleotide polymorphisms in case-control genome-wide association
studies. \textit{Biostatistics} \textbf{9} 201--215.

\bibitem[\protect\citeauthoryear{Arm}{2008b}]{2008bgail2} \textsc{Gail,
M. H.}, \textsc{Pfeiffer, R. M.}, \textsc{Wheeler, W.} and \textsc
{Pee, D}. (2008b).
Probability that a two-stage genome-wide association study will detect
a disease-associated SNP and implications for multistage designs. {\it
Ann. Hum. Genet.} \textbf{72} 812--820.


\bibitem[\protect\citeauthoryear{LIPTAK}{1958}]{1958LIPTAK} \textsc
{Liptak, T.} (1958). On the combination of independent tests. {\it
Magyar Tudomanyos Akademia Matematikai Kutato Intezetenek Kozlemenyei}
\textbf{3} 1971--1977.

\bibitem[\protect\citeauthoryear{Lou}{2004}]{2004lough} \textsc
{Loughin, T. M.} (2004). A systematic comparison of methods for
combining p-values from independent tests. \textit{Comput. Statist. Data
Anal.} \textbf{47} 467--485.
\MR{2086483}

\bibitem[\protect\citeauthoryear{Arm}{1959}]{1959mantel} \textsc
{Mantel, N.} and \textsc{Haenszel, W}. (1959). Statistical aspects of
the analysis of data from retrospective studies of disease. \textit{J.
Natl. Cancer Inst.} \textbf{22} 719--748.

%

\bibitem[\protect\citeauthoryear{Arm}{2003}]{2003pfei} \textsc
{Pfeiffer, R. M.} and \textsc{Gail, M. H.} (2003). Sample size
calculations for population- and family-based case-control association
studies on marker genotypes. \textit{Genet. Epidemiol.} \textbf{25} 136--148.

\bibitem[\protect\citeauthoryear{Arm}{1997}]{1997sas} \textsc{Sasieni,
P. D.} (1997). From genotypes to genes: Doubling the sample size. {\it
Biometrics } \textbf{53}
1253--1261.
\MR{1614374}

\bibitem[\protect\citeauthoryear{Arm}{2007}]{2007skol} \textsc{Skol,
A. D.}, \textsc{Scott, L. J.}, \textsc{Abacasis, G. R.} and \textsc
{Boehnke, M}. (2007). Optimal designs for two-stage genome-wide
association studies. \textit{Genet. Epidemiol.}
\textbf{31} 776--788.

\bibitem[\protect\citeauthoryear{Arm}{2006}]{2006skiol2} \textsc{Skol,
A. D.}, \textsc{Scott, L. J.}, \textsc{Abacasis, G. R.} and
\textsc{Boehnke, M}. (2006). Joint analysis is more efficient than
replication-based analysis for two-stage genome-wide association
studies. \textit{Nat. Genet.} \textbf{38} 209--213.

\bibitem[\protect\citeauthoryear{Arm}{1985}]{1985yusuf} \textsc{Yusuf,
S.}, \textsc{Peto, R.}, \textsc{Lewis, J.}, \textsc{Collins, R.}
and \textsc{Sleight, P.} (1985). Beta blockade during and after
myocardial infarction: An overview of the randomized trials. \textit{Prog.
Cardiovasc. Dis.} \textbf{27} 335--371.

\bibitem[\protect\citeauthoryear{Arm}{2007}]{2007yeager} \textsc
{Yeager, M.}, \textsc{Orr, N.}, \textsc{Hayes, R. B.} et al.
(2007). Genome-wide association study of prostate cancer identifies a
second risk locus at 8q24.
\textit{Nat. Genet.} \textbf{39} 645--649.

\bibitem[\protect\citeauthoryear{Arm}{2008}]{2008zegg} \textsc{Zeggini,
E.}, \textsc{Scott, L. J.}, \textsc{Saxena, R.} et al.
(2008). Meta-analysis of genome-wide association data and large-scale
replication identifies additional susceptibility loci for type 2
diabetes. \textit{Nat. Genet.} \textbf{40} 638--645.

\end{thebibliography}
\end{document}